# Asymmetric Nanoparticle May Go Active at Room Temperature

Nan Sheng[1], YuSong Tu[2], Pan Guo[1], RongZheng Wan[1], ZuoWei Wang[3,†], and HaiPing Fang[1,*]

[1] *Division of Interfacial Water and Key Laboratory of Interfacial Physics and Technology, Shanghai Institute of Applied Physics, Chinese Academy of Sciences, P.O. Box 800-204, Shanghai 201800, China*
[2] *College of Physics Science and Technology, Yangzhou University, Jiangsu, 225009, China*
[3] *School of Mathematical and Physical Sciences, University of Reading, Whiteknights, Reading RG6 6AX, United Kingdom*

**Using molecular dynamics simulations, we show that an asymmetrically shaped nanoparticle in dilute solution possesses a spontaneously curved trajectory within finite time interval, instead of the generally expected random walk. This unexpected dynamic behavior has a similarity to that of active matters, such as swimming bacteria, cells or even fishes, but is of a different physical origin. The key to the curved trajectory lies in the non-zero resultant force originated from the imbalance of the collision forces acted by surrounding solvent molecules on the shaped nanoparticle during its orientation regulation. Theoretical formulae based on the microscopic observation have been derived to describe this non-zero force and the resulted motion of the nanoparticle.**

The motion of molecules caused by thermal fluctuations plays an essential role in determining the probability for them to meet targets upon functioning [1–4], as found in many physical processes [3,5], chemical reactions [6,7] and biological functioning [2,3]. In conventional theories, the molecules/particles have been treated as perfect spheres with their trajectories described as random walks, following the original work of Einstein [8–11]. However, it has been realized by Einstein himself that this picture will break down if we can inspect the motion of the particles at much smaller time and length scales [9]. Significant progresses have been made on investigating the motion of particles at micrometers and timescales from microseconds to seconds, showing unconventional behaviors within relatively short time interval [12–18]. Han *et al.* experimentally observed a crossover from short-time anisotropic to long-time isotropic diffusion behavior of ellipsoidal particles along different axial directions [12]. Huang *et al.* experimentally measured the mean-square displacement of a 1-μm-diameter silica sphere in water using optical trapping technique and found that the particle motion cannot be described by conventional theory until sufficiently long time [14]. We note that, a majority of kinetic and dynamic processes related to molecules take place in nanoscale space [1–3,19–22] and accomplish in just several picoseconds [23,24], such as self-assembling [25–29], triggering chemical reaction [7,30], intercellular signal transduction [31], and neurotransmission [32]. Unfortunately, there is rare report on the unconventional behavior in the free motion of the molecules/nanoparticles solely under thermal fluctuations within short time at the nanoscale.

On the other hand, molecular dynamics (MD) simulation has been widely accepted as a powerful tool for studying the dynamics of molecules at nanoscales [33–42]. Our recent atomistic MD simulations showed interesting anisotropic motion of small asymmetric solute molecules, such as methanol and glycine, in water solely due to thermal fluctuations, which indicates the existence of rich dynamic behavior of asymmetric molecules in nano-space at finite timescales [43,44].

In this work, we report the emergence of an unexpected spontaneous curvature in the trajectory of an asymmetrically shaped nanoparticle in dilute solution, rather than random walk. This curvature results from a non-zero resultant force, which originates from the imbalance of the collision forces acted by surrounding solvent molecules on the shaped nanoparticle. The spontaneously curved trajectory of the shaped particle, together with the non-zero orientation-dependent force from surrounding solvent, is similar to the behavior of active matters, like swimming bacteria, cells or even fishes. However, different from the self-propelled active matter that consumes energy for generating driving force, the non-zero resultant force experienced by the shaped particle solely results from the collisions with the surrounding solvent during its orientation regulation under thermal fluctuations. Further, we derive theoretical formulae based on the observed microscopic picture that can well describe the non-zero force and the resulted motion of the nanoparticle.

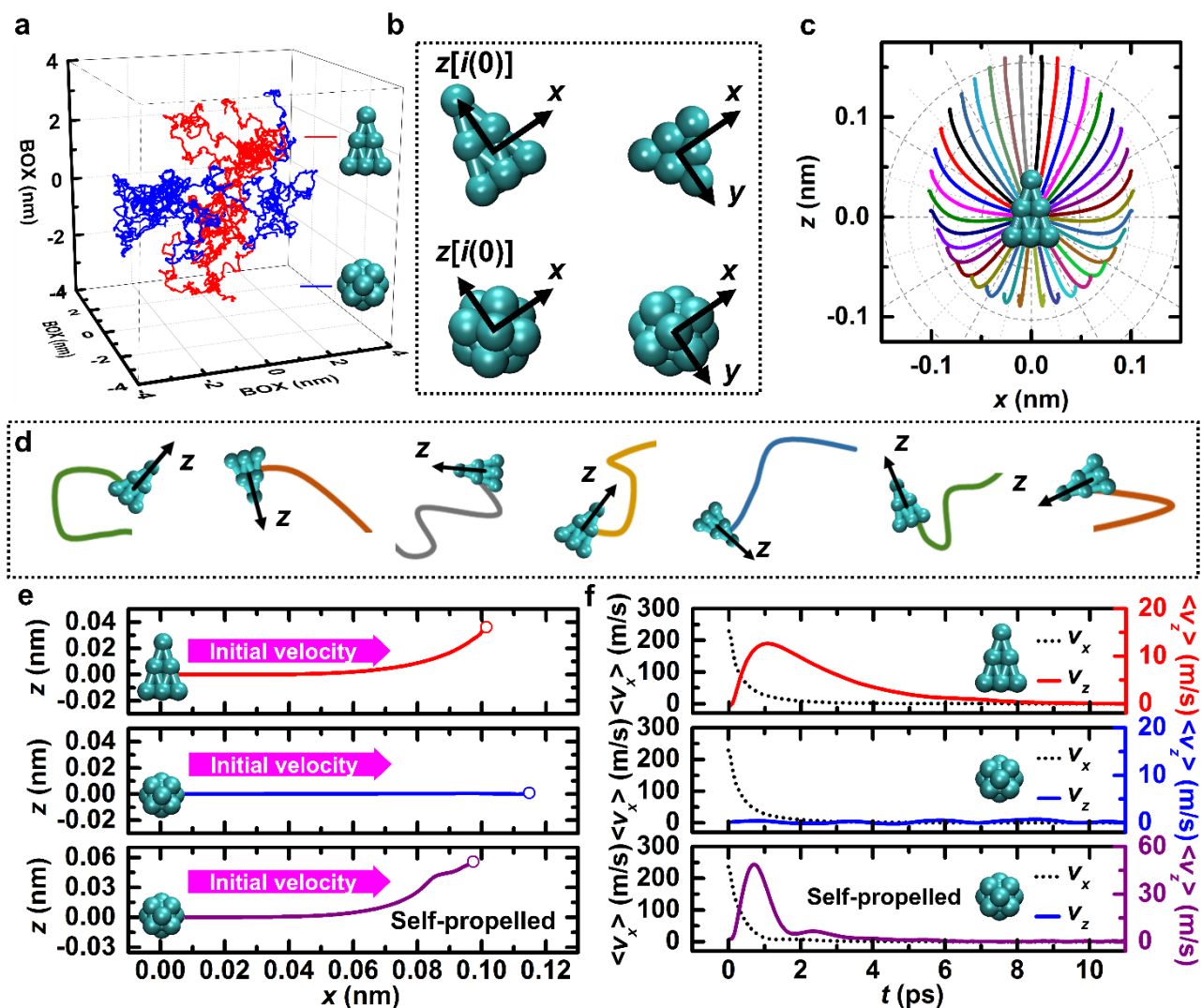

**Figure 1** (a) Typical trajectories of a pyramid-shaped nanoparticle with height of 0.37 nm and a spherical nanoparticle with diameter of 0.21 nm over 1 ns. (b) 3D Cartesian coordinate frames defined for the pyramid-shaped nanoparticle where the $z$-axis is along its *initial* orientation (from the center of the bottom face to the top atom) and the origin at its center of mass and for the spherical nanoparticle where the $z$-axis is along the *initial* vector pointing from its CoM to one atom on the surface. The left and right image on each row show the side and bottom views of each nanoparticle, respectively. Both frames are defined at initial time and fixed thereafter for data analysis. (c)

Ensemble-averaged trajectories (solid curves) of the pyramid-shaped nanoparticle in the *x-z* plane for different directions of the initial velocity as displayed in different colors. The dashed circles and lines are included for guiding the eyes. (d) Sampled 50ps trajectories of the nanoparticle starting from different **initial** orientations (*z*-axes). (e) Ensemble-averaged trajectories and (f) mean velocities of the two nanoparticles in the cases where their initial velocities are in direction along *x*-axis that is perpendicular to the initial orientations (*z*-axis), together with the simulation results of the self-propelled spherical nanoparticle as an active nanoparticle for comparison. The open circles in (d) mark the points where the mean CoM positions of the nanoparticles cease to move after about 10 ps.

**Results from MD simulations**

The simulation system consists of a single model nanoparticle shaped as triangular pyramid immersed in a solvent of small Lennard-Jones particles with periodic boundary conditions applied in all three directions, as described in the Simulation Method section in SM [45]. Each MD simulation was first run for 20 ns to equilibrate the system and then another 200 ns for data analysis. From five independent simulation runs, we collect abundant samples of the nanoparticle trajectories with the same time interval of 50 ps. As sketched in Fig 1(b,d), a three-dimensional (3D) Cartesian coordinate frame is defined for every sampled trajectory of the particle with the *z*-axis along its initial orientation (from the center of the bottom face to the top atom) and the origin at its center of mass (CoM). For comparison, we have also simulated another system where the shaped nanoparticle is replaced by a spherical nanoparticle with the same mass, atoms and volume whose orientation is defined by the vector pointing from its CoM to an atom on the surface.

Fig. 1(a) shows the typical trajectories of the pyramid-shaped and spherical nanoparticles over 1 ns, which demonstrate the homologous random feature of particle movement under thermal fluctuations. It is clear that the ensemble-averaged trajectory better reflects the intrinsic feature of the motion from noise. As shown in Fig. 1(e), just as one would generally expect without taking into account the shapes of molecules/particles, the averaged trajectory of the spherical nanoparticle follows a straight line along the direction of its initial velocity owing to the inertia. As the inertia decays owing to the collisions with the surrounding solvent, the averaged trajectory finally ends at a point. We note that the end point of the trajectory does not mean a motionless nanoparticle, but corresponds to an almost zero ensemble-averaged velocity owing to the isotropic probability of the motion. The pyramid-shaped nanoparticle shows a similar inertial motion at the initial stage. But surprisingly, as time increases, its ensemble-averaged trajectory spontaneously curves toward the direction of its original orientation (positive *z*-direction), although the initial velocity was in a perpendicular direction (positive *x*-direction). It is also interesting to note that in contrast to the monotonic decrease of the mean velocity, $<v_x(t)>$, of the shaped nanoparticle along its initial velocity direction (*x*-direction), its mean velocity along the *z*-direction, $<v_z(t)>$, rises up from zero to a peak value at $t = 1.08$ ps and then gradually declines to zero thereafter [Fig. 1(f)]. Accordingly, the mean CoM position of this nanoparticle drifts away from the origin and finally settles at a point with coordinates $x = 0.10$ nm and $z = 0.04$ nm after about 10 ps. Thus, the motion of the shaped nanoparticle comprises of two parts, i.e. the expected inertial motion along the initial velocity direction (here *x*-direction) and the unexpected directional motion towards the initial orientation direction (here *z*-direction). To provide a complete picture, Fig. 1(c) shows the ensemble-averaged trajectories of the shaped nanoparticle starting with initial velocities in all different directions (not only the *x*-direction as discussed above) with respect to the initial orientation (positive *z*-direction). Clearly, all of them display the bending curvature towards the *z*-direction.

Now we focus on understanding the physical origin of this spontaneously curved trajectory of the shaped nanoparticle by analyzing the ensemble-averaged force $<F_z^i(t)>$ acting on every individual constituent atom, where $i$ is the serial number of the atom under investigation. Apparently, $<F_z^i(t)>$ only results from the fluctuating forces owing to the collision with surrounding solvent molecules. We separate the contributions of these forces to $<F_z^i(t)>$ into two parts. The first part is independent of the velocities of the nanoparticle and its constituent atoms, which is equivalent to the effect of liquid pressure felt by a stationary object. As expected the ensemble average of such effects summing over all of the atoms is zero at the CoM of the particle, and therefore makes no contribution to the motion of the CoM. The other part is velocity-dependent and can be described in the form of the frictional force $<f_z^i(t)>$ given by the solvent against the motion of the atom. Following the Stoke's law, this force is considered to be linearly proportional to the mean velocity of the atom $<v_z^i(t)>$,

$$\langle f_z^i(t)\rangle = -\lambda^i \langle v_z^i(t)\rangle, \tag{1}$$

where $\lambda^i$ is the frictional coefficient of atom $i$. We note that the value of $\lambda^i$ varies for atoms located at different sites of the particle structure, depending on how they are in contact with the solvent. In practice, the frictional coefficient of each atom can be estimated from the linear relation between the measured mean frictional force $<f_z^i(t)>$ and mean velocity $<v_z^i(t)>$ (details shown in Sec. 1 of SM [45]).

The mean velocity of the $i$th atom along the $z$-direction, $<v_z^i(t)>$, can be written as

$$\langle v_z^i(t)\rangle = \langle v_z(t)\rangle - r_o^i(t) Q(t) \text{ with } Q(t) = -\frac{d}{dt} C_\varphi(t), \tag{2}$$

because the motion of each constituent atom could be separated into the CoM motion of the nanoparticle, denoted by $<v_z(t)>$ in Eq. (2), plus the rotation around the CoM denoted by the second term on the right-hand side (rhs) of Eq. (2). $r_o^i$ is the projection of the atom position vector $\mathbf{r}^i$, pointing from the particle CoM to the $i$th atom, on the particle orientation axis (unit vector denoted by $\mathbf{i}$), i. e., $r_o^i = \mathbf{r}^i \cdot \mathbf{i}$, and can be obtained directly from the construction of the nanoparticle structure. The differential $Q(t)$ of the particle orientation auto-correlation function, $C_\varphi(t) = <\mathbf{i}(0)\cdot\mathbf{i}(t)> = <\cos[\varphi(t)]>$ characterizes the particle rotational relaxation, where $\varphi(t)$ is the particle orientation angle with respect to its original orientation. Suppose that the shaped-particle rotates around the CoM with an angular velocity $\boldsymbol{\omega}$. As sketched in Fig. 2(a) for two typical constituent atoms, the velocity of atom $i$ has a rotation-related component of the form $\boldsymbol{\omega}\times\mathbf{r}^i$. When moving with this velocity, the atom experiences an effective frictional force due to the collision with surrounding molecules, $-\lambda^i(\boldsymbol{\omega}\times\mathbf{r}^i)$. Fig. 2(a) demonstrates that these effective forces differ for atoms located at different sites, which results in a non-zero net force at the CoM of the nanoparticle and consequently affects its movement. After taking ensemble average, the mean projection of the velocity of atom $i$ on the initial orientation $\mathbf{i}(0)$ is $<[\boldsymbol{\omega}(t)\times\mathbf{r}^i(t)]\cdot\mathbf{i}(0)> = -<[\mathbf{i}(0)\times\mathbf{r}^i(t)]\cdot\boldsymbol{\omega}(t)> = -r_o^i<[\mathbf{i}(0)\times\mathbf{i}(t)]\cdot\boldsymbol{\omega}(t)> = -r_o^i Q(t)$ and the mean effective frictional force is $\lambda^i r_o^i Q(t)$. We substitute Eq. (2) into Eq. (1) and sum up all the effective frictional forces acting on the constituent atoms of the shaped nanoparticle to get

$$\langle F_z(t)\rangle = \sum_{i=1}^{N}\langle f_z^i(t)\rangle = -mC_1\langle v_z(t)\rangle + mC_2 Q(t) \tag{3}$$

with

$$C_1 = \frac{\sum_{i=1}^{N}\lambda^i}{m}, \quad C_2 = \frac{\sum_{i=1}^{N}\lambda^i r_o^i}{m}. \tag{4}$$

where $m$ is the total mass of the nanoparticle. The first term on the rhs of Eq. (3) always resists the translational motion of the particle. It is the second term that could potentially provide an effective force generating the mean displacement of the particle CoM along the z-direction. We can calculate $C_1$ and $C_2$ directly based on Eq. (4) and obtain the values of $C_1 = 2.79$ ps$^{-1}$ and $C_2 = 0.11$ nm ps$^{-1}$ [detailed calculation shown in Sec. 1 of SM [45]). Then solving the Newton's second law which is a differential equation of velocity, we can obtain the mean velocity $\langle v_z(t)\rangle$,

$$\langle v_z(t)\rangle = C_2 R(t) \text{ with } R(t) = e^{-C_1 t}\int_0^t Q(\xi)e^{C_1\xi}d\xi. \quad (5)$$

We have also computed the auto-correlation function $C_\varphi(t)$ of the nanoparticle orientation from the MD trajectories. In classical rotational Brownian motion, $C_\varphi(t)$ is predicted to decay exponentially with time [11]. However, our simulation data in Fig. 2(b) show that in the very first picosecond where the inertial effect dominates, $C_\varphi(t)$ does not follow the exponential form. A standard bi-exponential function of the form

$$C_\varphi(t) = \frac{\tau_1}{\tau_1-\tau_2}e^{-t/\tau_1} - \frac{\tau_2}{\tau_1-\tau_2}e^{-t/\tau_2}, \quad (6)$$

with the characteristic times $\tau_1 = 1.82$ ps and $\tau_2 = 0.25$ ps is found to describe the data very well.

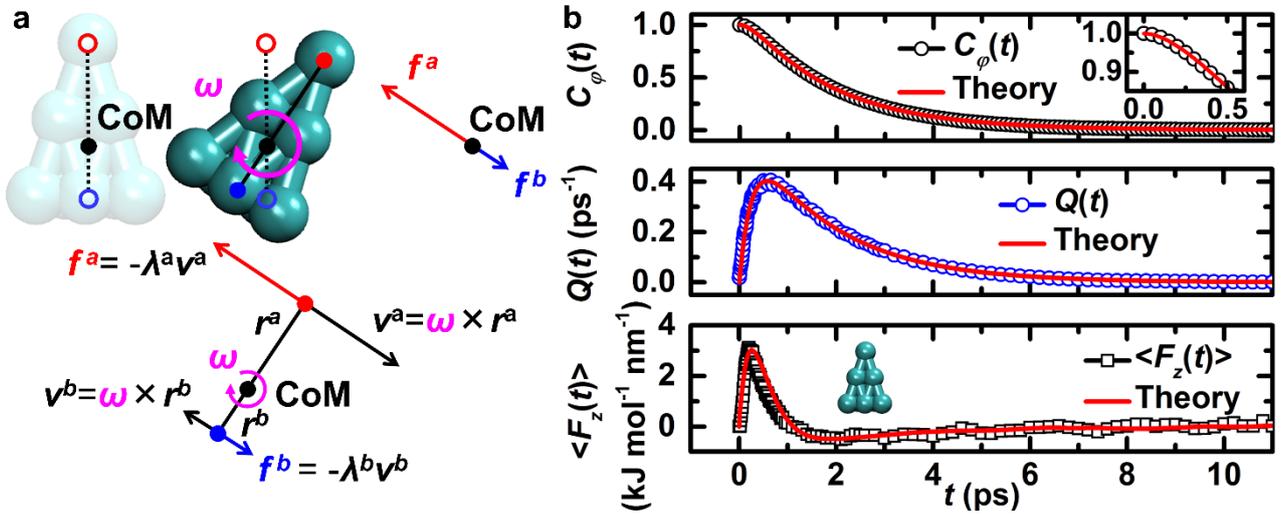

**Figure 2** (a) Sketch of the rotation-related velocities of one top and one bottom atoms of the pyramid-shaped nanoparticle and the effective frictional forces acting on them by surrounding solvent molecules. The contributions of these forces on determining the translational motion of the particle via its center of mass are shown in the lower part of the plot. (b) Autocorrelation function of the particle orientation $C_\varphi(t)$ (black circles), its rate of variation $Q(t)$ (blue circles) and the mean force experienced by the CoM of the nanoparticle along the z-axis (black squares). The inset is an enlarged image of $C_\varphi(t)$ in the very first half picosecond. The red curves are theoretical predictions of Eq. (6), its derivative and Eq. (7), respectively.

The mean force acting on the CoM of the nanoparticle is thus obtained by substituting Eqs. (5,6) into Eq. (3)

$$\langle F_z(t)\rangle = mC_2[Q(t) - C_1 R(t)], \quad (7)$$

which provides good description of the simulation results as shown in Fig. 2(b). It is clear that $\langle F_z(t)\rangle$ does have a positive value towards the initial orientation **i**(0) of the particle at early time. More specifically this force first increases from zero, reaches its maximal value at about 0.2 ps and then decreases. It becomes negative at 1 ps, reaches the minimal value at about 2 ps and then gradually approaches zero. Eq. (7) further clarifies that the mean force on the nanoparticle can be nonzero only when $C_2 \neq 0$ and $[Q(t) - C_1 R(t)] \neq 0$. The contribution of the asymmetric architecture of the nanoparticle is carried by $C_2$, since it possesses the position vector of every atom. The terms in the square bracket are all rotational motion related. It indicates that the key reason for the non-zero net force $<F_z(t)>$ lies in the imbalance of the effective frictional forces or hydrodynamic resistances ($C_2 \neq 0$) acting on the constituent atoms during the particle orientation regulation. This can only happen for particles with asymmetric structures. If the geometric shape of the particle is symmetric, the effective frictional forces acting on all atoms are well balanced ($C_2 = 0$) and no directional motion can be observed, see examples given in SM (Sec. 2) [45]. On the other hand, the rotational motion of the nanoparticle also plays an important role. If we fix the orientation of the particle [$C_\varphi(t) = 1$], $Q(t)$ and $R(t)$ in Eq. (7) both equal to zero. Then even for an asymmetrically shaped nanoparticle with $C_2 \neq 0$ the mean force is zero. Therefore, the key of nonzero mean force is asymmetry, while the rotation is also required.

Based on the theoretical analyses above, we impose *an additional force* of $\mathbf{F}_{add}(t) = 340[\mathbf{i}(t) \times \mathbf{L}(t)]$ pN on the center of mass of *a spherical nanoparticle*, where $\mathbf{i}(t)$ is the unit vector pointing from the particle CoM to a certain atom on the surface and $\mathbf{L}(t)$ is the angular momentum of the rotation. This self-propelled active nanoparticle is found to possess a very similar curved mean trajectory to that of the pyramid-shaped nanoparticle. As shown in Fig.1(e), it also bends towards the direction of its original orientation **i** (positive *z*-direction), although the initial velocity was in a perpendicular direction (positive *x*-direction). It settles at a final position with coordinates $x = 0.10$ nm and $z = 0.05$ nm. The average velocity of the self-propelled particle along the *z*-direction rises from zero to the peak at $t = 0.71$ ps and then declines to zero thereafter. There is a small bulge in the velocity at $t = 2.33$ ps causing a small wave near the tail of the trajectory, which might be due to some hidden influence of the additional force during the rotational relaxation. The similarity in the curved trajectories of the self-propelled spherical nanoparticle and the freely moving pyramid-shaped nanoparticle suggests that the shaped nanoparticle bears certain extent of activity analogous to self-propelled active matter.

To summarize, we have shown by molecular dynamics simulations that the motion of a shaped nanoparticle with broken central symmetry in dilute solution possesses spontaneously curved trajectory within a finite time interval, instead of the generally expected random walk. This unexpected dynamic behavior has a similarity to that of active matters [49–52]. However, different from the self-propelled active matter which consumes energy to provide driving force [52–55], here the non-zero resultant force solely results from the thermal fluctuations. We have also derived theoretical formulae that can well describe the physical origin of this non-zero resultant force.

A further remark we would like to make is that the observed curvature in the trajectory will not lead to a perpetual mobile that violates the second law of thermodynamics. As shown above, the trajectory spontaneously bends towards the direction along the initial orientation of the shaped particle. In equilibrium systems with sufficiently long time or sufficiently large number of particles for averaging, the orientations of the nanoparticles have equal probability in all directions. After averaging over all possible initial orientations of the particles, the mean displacement is zero and so there is no directional flow in the system, which is consistent with the statistical mechanic principles. In the SM (Sec. 5) [45], we have demonstrated that one cannot extract mechanical energy from a thermal bath by restraining the orientations of the particles. However, for the case that we only focus on the motion of a single molecule within a finite time, we will see the directional curved trajectory and the spontaneous active motion.

Considering that a majority of physical, chemical and biological processes happen within nano-space and within finite timescales and most of the involved molecules possess asymmetric structures, we expect that the directional curved trajectory may have significant influence on various processes such as nucleation of clusters, self-assembling, triggering chemical reaction, intercellular signal transduction, neurotransmission. For example, the intracellular signaling is usually carried out by small signal molecules with distance of several nanometers to their receptors and the time is very finite [31,56]. However, how to evaluate the role of the directional curved trajectory in the dynamics of these processes still remain to be an open and challenging research subject.

We acknowledge Alexei Likhtman, Alex Lukyanov, Eugene Terentjev and Qing JI for valuable discussions. This work was supported by the NNNSFC (10825520, 11422542, 11175230 and 11290164), the KRPCAS (KJZD-EW-M03), the Supercomputing Center of Chinese Academy of Sciences and the Shanghai Supercomputer Center of China.

* Corresponding author. Email: fanghaiping@sinap.ac.cn
† Corresponding author. Email: zuowei.wang@reading.ac.uk

**References**
[1]   P. Ball, Chem. Rev. **108**, 74 (2008).
[2]   B. L. de Groot and H. Grubmüller, Science **294**, 2353 (2001).
[3]   Y. von Hansen, S. Gekle, and R. R. Netz, Phys. Rev. Lett. **111**, 118103 (2013).
[4]   P. Král, L. Vukovi, N. Patra, B. Wang, K. Sint, and A. Titov, J. Nanosci. Lett. **1**, 1332 (2011).
[5]   J. H. Bahng, B. Yeom, Y. Wang, S. O. Tung, J. D. Hoff, and N. Kotov, Nature **517**, 596 (2015).
[6]   P. Hervés, M. Pérez-Lorenzo, L. M. Liz-Marzán, J. Dzubiella, Y. Lu, and M. Ballauff, Chem. Soc. Rev. **41**, 5577 (2012).
[7]   J. Yang, G. Shi, Y. Tu, and H. Fang, Angew. Chemie - Int. Ed. **53**, 10190 (2014).
[8]   A. Einstein, Ann. Phys. **322**, 549 (1905).
[9]   A. Einstein, Zeitschrift Für Elektrotechnik Und Elektrochemie **13**, 41 (1907).
[10]  D. Chandler, *Introduction to Modern Statistical Mechanics* (Oxford University Press, 1987).
[11]  M. Doi and S. F. Edwards, *The Theory of Polymer Dynamics* (Clarendon Press, 1988).
[12]  Y. Han, A. M. Alsayed, M. Nobili, J. Zhang, T. C. Lubensky, and A. G. Yodh, Science **314**, 626 (2006).
[13]  T. Li and M. G. Raizen, Ann. Phys. **525**, 281 (2013).
[14]  R. Huang, I. Chavez, K. M. Taute, B. Lukić, S. Jeney, M. G. Raizen, and E.-L. Florin, Nat. Phys. **7**, 576 (2011).
[15]  A. Chakrabarty, A. Konya, F. Wang, J. V. Selinger, K. Sun, and Q.-H. Wei, Phys. Rev. Lett. **111**, 160603 (2013).
[16]  J. Zhang, X. Xu, and T. Qian, Phys. Rev. E **91**, 033016 (2015).
[17]  H. Brenner, J. Colloid Sci. **20**, 104 (1965).
[18]  W. A. Wegener, Biopolymers **20**, 303 (1981).
[19]  A. Schlaich, E. W. Knapp, and R. R. Netz, Phys. Rev. Lett. **117**, 048001 (2016).
[20]  H. Qiu, X. C. Zeng, and W. Guo, ACS Nano **9**, 9877 (2015).
[21]  L. Ma, A. Gaisinskaya-Kipnis, N. Kampf, and J. Klein, Nat. Commun. **6**, 6060 (2015).
[22]  A. Barati Farimani, N. R. Aluru, and E. Tajkhorshid, Appl. Phys. Lett. **105**, 083702 (2014).
[23]  T. Franosch, M. Grimm, M. Belushkin, F. M. Mor, G. Foffi, L. Forró, and S. Jeney, Nature **478**, 85 (2011).
[24]  P. N. Pusey, Science **332**, 802 (2011).
[25]  D. Chandler, Nature **437**, 640 (2005).

[26] L. Zhao, C. Wang, J. Liu, B. Wen, Y. Tu, Z. Wang, and H. Fang, Phys. Rev. Lett. **112**, 078301 (2014).
[27] N. Arai, K. Yasuoka, and X. C. Zeng, J. Am. Chem. Soc. **130**, 7916 (2008).
[28] A. Reinhardt and D. Frenkel, Soft Matter **12**, 6253 (2016).
[29] X. Zhou, G. Liu, K. Yamato, Y. Shen, R. Cheng, X. Wei, W. Bai, Y. Gao, H. Li, Y. Liu, F. Liu, D. M. Czajkowsky, J. Wang, M. J. Dabney, Z. Cai, J. Hu, F. V. Bright, L. He, X. C. Zeng, Z. Shao, and B. Gong, Nat. Commun. **3**, 949 (2012).
[30] T. Guérin, O. Bénichou, and R. Voituriez, Nat. Chem. **4**, 568 (2012).
[31] M. J. Berridge and R. F. Irvine, Nature **341**, 197 (1989).
[32] B. Barbour and M. Häusser, Trends Neurosci. **20**, 377 (1997).
[33] R. Wan, C. Wang, X. Lei, G. Zhou, and H. Fang, Phys. Rev. Lett. **115**, 195901 (2015).
[34] P. Guo, Y. Tu, J. Yang, C. Wang, N. Sheng, and H. Fang, Phys. Rev. Lett. **115**, 186101 (2015).
[35] Y. Huang, C. Zhu, L. Wang, X. Cao, Y. Su, X. Jiang, S. Meng, J. Zhao, and X. C. Zeng, Sci. Adv. **2**, e1501010 (2016).
[36] G. Hummer, J. C. Rasaiah, and J. P. Noworyta, Nature **414**, 188 (2001).
[37] J. Yang, S. Meng, L. F. Xu, and E. G. Wang, Phys. Rev. Lett. **92**, 146102 (2004).
[38] B. Wang and P. Král, Phys. Rev. Lett. **101**, 046103 (2008).
[39] I. Kosztin and K. Schulten, Phys. Rev. Lett. **93**, 238102 (2004).
[40] C. Zhu, H. Li, Y. Huang, X. C. Zeng, and S. Meng, Phys. Rev. Lett. **110**, 126101 (2013).
[41] R. Zhou, X. Huang, C. J. Margulis and B. J. Berne, Science **305**, 1605 (2004).
[42] Q.-L. Zhang, W.-Z. Jiang, J. Liu, R.-D. Miao, and N. Sheng, Phys. Rev. Lett. **110**, 254501 (2013).
[43] N. Sheng, Y. Tu, P. Guo, R. Wan, and H. Fang, J. Hydrodyn. Ser. B **24**, 969 (2012).
[44] N. Sheng, Y. Tu, P. Guo, R. Wan, and H. Fang, Sci. China Physics, Mech. Astron. **56**, 1047 (2013).
[45] See Supplemental Material for simulation method, more detailed simulation data and additional tests, which also includes Refs. [45–47].
[46] I.-C. Yeh and G. Hummer, J. Phys. Chem. B **108**, 15873 (2004).
[47] G. Bussi, D. Donadio, and M. Parrinello, J. Chem. Phys. **126**, 014101 (2007).
[48] B. Hess, C. Kutzner, D. van der Spoel, and E. Lindahl, J. Chem. Theory Comput. **4**, 435 (2008).
[49] M. C. Marchetti, J. F. Joanny, S. Ramaswamy, T. B. Liverpool, J. Prost, M. Rao, and R. A. Simha, Rev. Mod. Phys. **85**, 1143 (2013).
[50] C. Krüger, G. Klös, C. Bahr, and C. C. Maass, Phys. Rev. Lett. **117**, 048003 (2016).
[51] P. K. Ghosh, Y. Li, G. Marchegiani, and F. Marchesoni, J. Chem. Phys. **143**, 211101 (2015).
[52] É. Fodor, C. Nardini, M. E. Cates, J. Tailleur, P. Visco, and F. van Wijland, Phys. Rev. Lett. **117**, 038103 (2016).
[53] P. Romanczuk and L. Schimansky-Geier, Phys. Rev. Lett. **106**, 230601 (2011).
[54] I. Buttinoni, G. Volpe, F. Kümmel, G. Volpe, and C. Bechinger, J. Phys. Condens. Matter **24**, 284129 (2012).
[55] H. Wioland, F. G. Woodhouse, J. Dunkel, J. O. Kessler, and R. E. Goldstein, Phys. Rev. Lett. **110**, 268102 (2013).
[56] N. A. Campbell and J. B. Reece, *Biology* (Benjamin Cummings, 2001).